\documentclass[12pt]{article}

\usepackage{graphicx}  
\usepackage{dcolumn}   
\usepackage{bm}        
\usepackage{amssymb}   
\usepackage{amsmath}
\usepackage{color}
\usepackage[colorlinks=true, pdfpagemode=UseNone, pdfstartview=FitH, linkcolor=blue, citecolor=blue, urlcolor=blue]{hyperref} 
\usepackage{epstopdf}
\usepackage{dsfont}

\usepackage{scicite}

\usepackage{times}

\usepackage[english]{babel}
\usepackage[utf8]{inputenc}
\usepackage[T1]{fontenc}

\newenvironment{sciabstract}{%
\begin{quote} \bf}
{\end{quote}}

\newcounter{lastnote}

\title{Classical many-body time crystals\\\vspace{0.5cm} \normalsize{Time crystals are readily obtained in the steady state of many-body classical systems that undergo period-doubling bifurcations.}}

\author{Toni L. Heugel$^{1,\ast}$, Matthias Oscity$^{1,2,\ast}$, Alexander Eichler$^{3}$, Oded Zilberberg$^{2}$, \\ and R. Chitra$^{1}$\\
\\
\normalsize{$^1$ Institute for Theoretical Physics, ETH Z\"{u}rich, Wolfgang-Pauli-Stra\ss e 27, 8093 Z\"urich, Switzerland.}\\
\normalsize{$^2$ Fachhochschule Nordwestschweiz FHNW, Klosterzelgstrasse 2, CH-5210 Windisch, Switzerland.}\\
\normalsize{$^3$ Institute for Solid State Physics, ETH Z\"{u}rich, Wolfgang-Pauli-Stra\ss e 27, 8093 Z\"urich, Switzerland.}\\
\normalsize{$^\ast$ These authors contributed equally.}\\
}

\date{}

\begin{document} 


\baselineskip24pt


\maketitle


\begin{sciabstract}
Discrete time crystals are a many-body state of matter where the extensive system's dynamics are slower than the forces acting on it. Nowadays, there is a growing debate regarding the specific properties required to demonstrate such a many-body state, alongside several experimental realizations. In this work, we provide a simple and pedagogical framework by which to obtain many-body time crystals using parametrically coupled resonators. In our analysis, we use classical period-doubling bifurcation theory and present a clear distinction between single-mode time-translation symmetry breaking and a situation where an extensive number of degrees of freedom undergo the transition. We experimentally demonstrate this paradigm using coupled mechanical oscillators, thus providing a clear route for time crystals realizations in real materials.
\end{sciabstract}

In periodically modulated nonlinear systems, discrete time-translation symmetry can be spontaneously broken, leading to inherently slower dynamics than that of the drive~\cite{Faraday1831, Mathieu1868, Rayleigh1887, Landau_Lifshitz_Mechanics, Nayfeh2007, Dykmanbook}. A rapidly expanding community is principally focused on such a phenomenon in periodically-driven closed quantum systems, where disorder and interactions are considered to be essential for so-called discrete time crystals~\cite{Khemani2016, Else2016,Else2017, Keyserlingk2016, Keyserlingk2016A,Yao2017, Zhang2017,  Choi2017,Ho2017,Sacha2017,Dykman2018,  Rovny2018,Berdanier2018, Yao2018}. A time-crystalline phase of matter stabilized by many-body localization was first observed in a one-dimensional trapped-ion system~\cite{Zhang2017}. Surprisingly, time crystals were also seen in three-dimensional ensembles of NV-centers~\cite{Choi2017} and in spin-$1/2$ nuclei in phosphate materials~\cite{Rovny2018A} where disorder-induced localization effects are absent. The latter results indicate a wider class of time-crystalline behavior, including classical counterparts~\cite{Yao2018A}.

A natural arena for realizing time crystals is provided by parametric resonators. A parametrically-pumped resonator mode plays an important role in many areas of science and technology. In its best-known form, parametric pumping describes the modulation of a resonator's potential at twice its natural frequency~\cite{Faraday1831, Mathieu1868, Rayleigh1887, Landau_Lifshitz_Mechanics}. When the modulation depth exceeds an instability threshold, the resonator undergoes a period-doubling bifurcation to a new regime stabilized by nonlinearities~\cite{Landau_Lifshitz_Mechanics}. This time-translation symmetry breaking (TTSB) leads to two stable \textit{parametric phase states} that have equal amplitude, opposite phase, and half the oscillation frequency of the parametric drive~\cite{Papariello2016,Leuch2016, Eichler2018, Dykman2018}. Interestingly, these states can be associated with two states of a classical bit~\cite{Goto1959, Woo1971, Mahboob2008, Mahboob2011, Mahboob2014} or with an Ising spin~\cite{Goto2016,Nigg2017, Puri2017, Goto2018,Wang2013,McMahon2016,Inagaki2016,Inagaki2016a, Ryvkine2006, Chan2008, Lin2015, Papariello2016, Leuch2016, Eichler2018}. Network of such coupled resonators have been proposed as simulation platforms for complex Ising-like models that are very hard to solve with conventional computers~\cite{Dykman2018,Goto2016, Nigg2017, Puri2017,Goto2018, Wang2013,McMahon2016,Inagaki2016,Inagaki2016a}.


In this work, we show that a many-body TTSB can be easily realized in a classical network of dissipative parametric resonators. We present a general theoretical analysis and derive conditions for the manifestation of {\it many-body} TTSB in this system. This is complemented by a simple tabletop experimental demonstration using two coupled resonators. Our setup allows us to tune the coupling strength, and we find a regime where the modes of the system jointly undergo TTSB into well-defined parametric phase state configurations. Our experiment thus realizes the simplest building block that highlights the plethora of accessible TTSB solutions. At the same time, we test our understanding of the general many-body model against a well-controlled and accessible experimental implementation. Our work lifts the ambiguity surrounding the concept of time crystals by establishing sufficient conditions for their generation.

We consider a classical network of $N$ coupled nonlinear parametric oscillators, whose dynamics is governed by $N$ equations of motion
\begin{align}
\ddot{x}_i + \omega_{i}^2 [1 &- \lambda \cos(\Omega_p t)] x_i + \gamma_i \dot{x}_i \nonumber\\&+ \alpha_i x_i^3 + \eta_i x_i^2 \dot{x}_i  -\sum_{i\ne j} \beta_{ij}x_j= 0 \,, 
\label{many}
\end{align}
where dots mark differentiations with respect to time $t$, $x_i$ is the displacement, $\omega_{i}$ is the eigenfrequency, $\gamma_i$ the dissipation, $\alpha_i$ the quartic nonlinearity, and $\eta_i$ the nonlinear damping of the $i^{\rm th}$ mode. The system is excited by a single parametric pump of modulation depth $\lambda$ and  frequency $\Omega_p$. Each mode $i$ couples to other modes $j \neq i$ in the form of a driving force in proportion to $x_i$ and with a coupling coefficient $\beta_{ij}$.

 We can perturbatively solve the system using the slow-flow method~\cite{Guckenheimer1983}:  we rewrite Eq.~\eqref{many} as $2N$ first-order differential equations and perform a van der Pol transformation with frequency $\omega=\Omega_p/2$, followed by time-averaging, to obtain the slow-flow equation
	\begin{equation}
		\dot{\bm{X}} = A(\bm{X}) \bm{X}\,,
		\label{eq:coupled_slow_7}
	\end{equation}
where $\bm{X} = (u_1,v_1,u_2,v_2,\ldots, u_N,v_N)^T$, with $u_i$ and $v_i$ the slowly varying phase-space quadratures of the individual resonators. 
This equation is valid if the dimensionless quantities $1-(\frac{\omega}{\omega_i})^2$, $\lambda$, $\gamma_i/\omega_{i}$, $\frac{\eta_i }{\omega_{i}} x_i^2$, $\frac{\beta_{ij}}{\omega_{i}^2}$, and $\frac{\alpha_i}{\omega_{i}^2} x_i^2$ are of order $\epsilon$, where $0<\epsilon\ll1$~\cite{Guckenheimer1983}.  These conditions are
easily satisfied for a network of nearly identical oscillators.
 The matrix $A$ can be written as 
\begin{equation}
	A= \begin{pmatrix}
a_1(\bm{X}) & b_{12} & \cdots &b_{1N}\\
b_{1,2}& a_2(\bm{X}) & \ddots  & \vdots \\
\vdots & \ddots &\ddots &b_{(N-1)N} \\
b_{1N}&\cdots & b_{(N-1)N} & a_N(\bm{X})
\end{pmatrix},
\end{equation}
where the $a_i$ and $b_{ij}$ are given by, 
\[
	a_i(\bm{X})  =-\frac{1}{8\omega} \begin{pmatrix}
	  a_{i,1} & a_{i,2} \\  
	a_{i,3}  & a_{i,4} \end{pmatrix}, \quad
	b_{ij} = \begin{pmatrix}
	0 &  \frac{\beta_{ij}}{2 \omega} \\  
	-\frac{\beta_{ij}}{2 \omega} & 0 
	\end{pmatrix},
\]
with $i\ne j$ and $i,j=1,2,\ldots,N$, and using the definitions $a_{i,1}=a_{i,4}=4\gamma_i\omega + \eta_i\omega X_i^2$, $a_{i,2}=2 \left(\lambda \omega_{i}^2 +2(\omega_{i}^2 - \omega^2)\right) + 3 \alpha_i X_i^2$, $a_{i,3}=2 \left(\lambda \omega_{i}^2 -2(\omega_{i}^2 - \omega^2)\right) -  3 \alpha_i X_i^2$, and $X_i = u_i^2 + v_i^2$. 
In general, the number of steady-state solutions, both stable and unstable,  to this $N$-body problem varies from $1$ to $5^N$ depending on the parameter regimes~\cite{Papariello2016}.

\begin{figure}[ht!]
\includegraphics[width=0.5\linewidth]{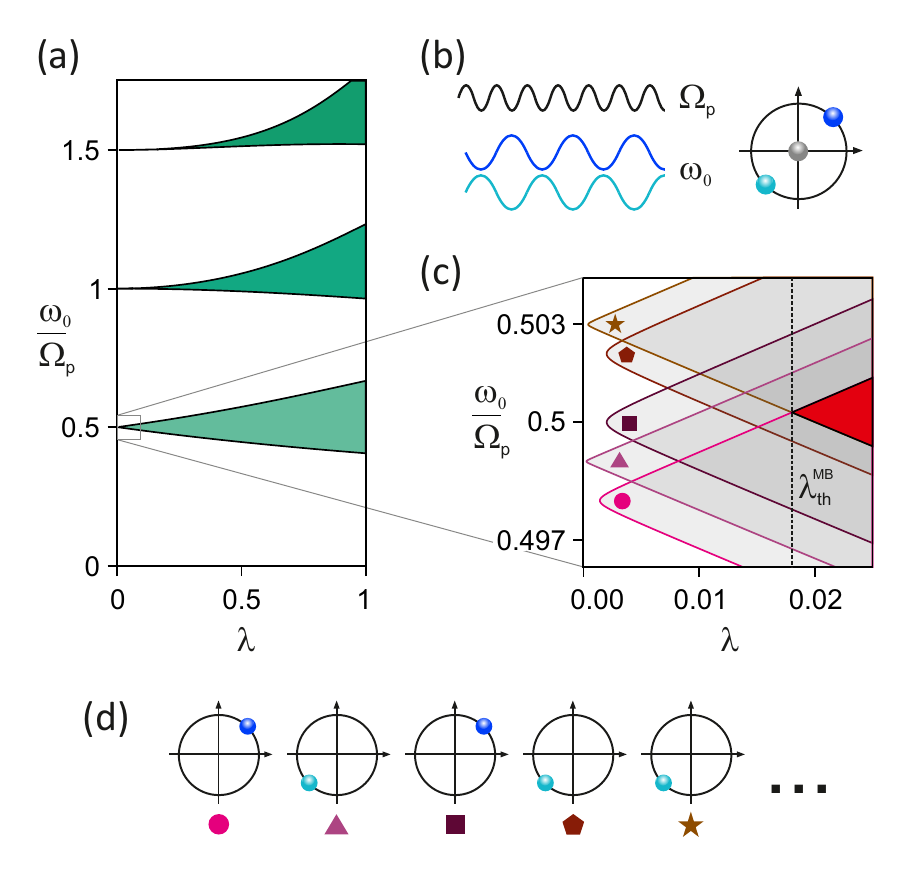}
\caption{\label{fig:figure1} (a) Stability diagram of a single parametrically-driven resonator mode with eigenfrequency $\omega_0$ as a function of parametric pumping frequency and depth, $\Omega_p$ and $\lambda$, respectively~\cite{Landau_Lifshitz_Mechanics, Hanggi1994}, cf.~Eq.~\eqref{many}. Green tongue shapes indicate regions where the linear resonator becomes unstable. (b) Beyond the instability threshold in the first lobe, $\Omega_p\sim 2 \omega_0$, the nonlinear resonator undergoes a period-doubling bifurcation and oscillates with frequency $\Omega_p/2$. Due to the period doubling, there exist two possible phase states with equal amplitude but opposite phase. In some frequency ranges, there is an additional zero-displacement stable solution. (c) Zoom of the stability diagram around the first instability lobe for $N$ coupled resonators. Here, $N$ nondegenerate normal modes (marked by different color and symbols) generically arise, cf.~Eqs.~\eqref{many}. The red area indicates the region of many-body TTSB. (d) Inside the many-body TTSB, each of the normal modes resides in one of the phase states. The resulting multi-state configuration depends on the coupling coefficients $\beta_{ij}$, the nonlinearities, and noise fluctuations.}
 \end{figure}

In the absence of nonlinearities, $\alpha_i=\eta_i=0$, the natural description of the resonator network is given by $N$ normal modes with eigenfrequencies $\nu_{k}$, $k=1,\ldots,N$. The dynamics of the normal modes is determined by the eigenvalues and the eigenvectors of $A$. The $2N$ eigenvectors define the positions and momenta of the $N$ normal modes.  The time evolution of the $k$-th normal mode is given by $e^{\Lambda_k t}$, with $\Lambda_k$ the respective eigenvalue. The motion will be bounded for negative ${\rm Re}\{\Lambda_k\}$ and manifest parametric instability, i.e., unbounded dynamics when ${\rm Re}\{\Lambda_k\} >0$. Each normal mode exhibits a  corresponding parametric stability phase diagram known as `Arnold tongues', delineating regions where dissipation stabilizes the motion and regions where the linear  system shows unbounded dynamics, see Fig.~\ref{fig:figure1}(a). In the following, we will focus on the dominant instability lobe occurring around twice the natural frequency of the normal mode $k$, $\Omega_p\sim 2\nu_{k}$, when the parametric drive exceeds a threshold $\lambda\geq\lambda_{\rm th}^k$
~\cite{Hanggi1994}. 


In general, it is not dissipation but the underlying nonlinearities $(\alpha_i,\eta_i)$ that stabilize the normal-mode oscillations against unbounded growth~\cite{Nayfeh2007}. At the boundary of its main instability lobe, each normal mode undergoes a period-doubling bifurcation alongside a spontaneous $Z_2$ symmetry breaking between the two parametric phase states, see Fig.~\ref{fig:figure1}(b). This is a simple manifestation of TTSB in the steady state of an effective single parametric mode. The parametric phase states define an effective Ising-like phase bit~\cite{Goto1959, Woo1971, Mahboob2008, Mahboob2011, Mahboob2014}. It is important to note that although a single normal mode can involve an extensive number of resonators of the network, it does not give rise to a many-body TTSB because it does not involve an extensive number of independent degrees of freedom.

A many-body TTSB phase is realized in the resonator network in a region where an extensive number of normal modes undergo  the aforementioned period-doubling transition. A simple recipe to realize a many-body TTSB consists of finding the parametric pumping amplitude $\lambda_{\rm th}^{\rm MB}(\omega_P)=\min_{\lambda} \{\lambda > \lambda_{\rm th}^k(\omega_P)\,,\,\forall k\}$ at which all normal modes are driven above their respective instability thresholds, see Fig.~\ref{fig:figure1}(c).  There, each normal mode finds itself in a parametric phase state, see Fig.~\ref{fig:figure1}(d). Note that the many-body threshold holds in the limit of weak nonlinearities and does not include corrections  stemming from nonlinear inter-normal mode coupling.
In  the mean-field  limit of  $N$ identical resonators, i.e.,  $\omega_i\equiv\omega_0$ and  $\gamma_i=\gamma$, with all-to-all coupling $\beta_{ij}= \beta/\sqrt{N}$,   apart from the symmetric mode, all other   instability lobes  coincide with that of the     antisymmetric ($a$)   mode. The  respective instability thresholds  ($\lambda>\lambda_{\rm th})$ are given by~\cite{supmat}:
\begin{equation}
\label{threshMB}
	 \lambda_{\rm th}^{s/a}  = \frac{4\omega}{\omega_0 ^2}\sqrt{ \frac{\gamma^2}{4} +  \left( \frac{\omega^2 - \omega_0^2}{2\omega } + 
		\left\{\begin{array}{lr} 
		\frac{(N-1)}N,& \text{s}  \\
		\frac{-1}N,& \text{a}
		\end{array}\right\}
		\frac{\beta^2}{2 \omega}\right)^2}.
\end{equation}
 The overlap region of $\lambda_{\rm th}^{s/a}$ defines  $\lambda\geq\lambda_{\rm th}^{\rm MB}(\Omega_p)$.

In the following we discuss two limits, `strong' and `weak' coupling, that  are defined relative to the parametric modulation strength, $\lambda$. For weak  $\beta_{ij}$,  the normal modes closely resemble the underlying constituent resonators.  However, as  $\beta_{ij}$ increase, the normal modes  become collective in nature.  In the many-body TTSB phase the system can choose one of $2^N$ to $3^N$ configurations: in the weak coupling regime, these correspond  to the possible configurations of the $N$ individual resonators~\cite{Dykman2018,Danzl2010}, while in the strong coupling regime, they correspond to the configurations of collective normal modes.  In both cases,  all  these configurations manifest TTSB and the chosen configuration will depend on initial conditions, noise and the strength of the nonlinearities.  To summarize, we predict that an array of coupled  dissipative parametric resonators realizes a  stable TTSB phase in its steady state. This  phase  endures in a wide region of parameter space and is robust to fluctuations.  

\begin{figure*}[t!]
\includegraphics[width = 1\textwidth]{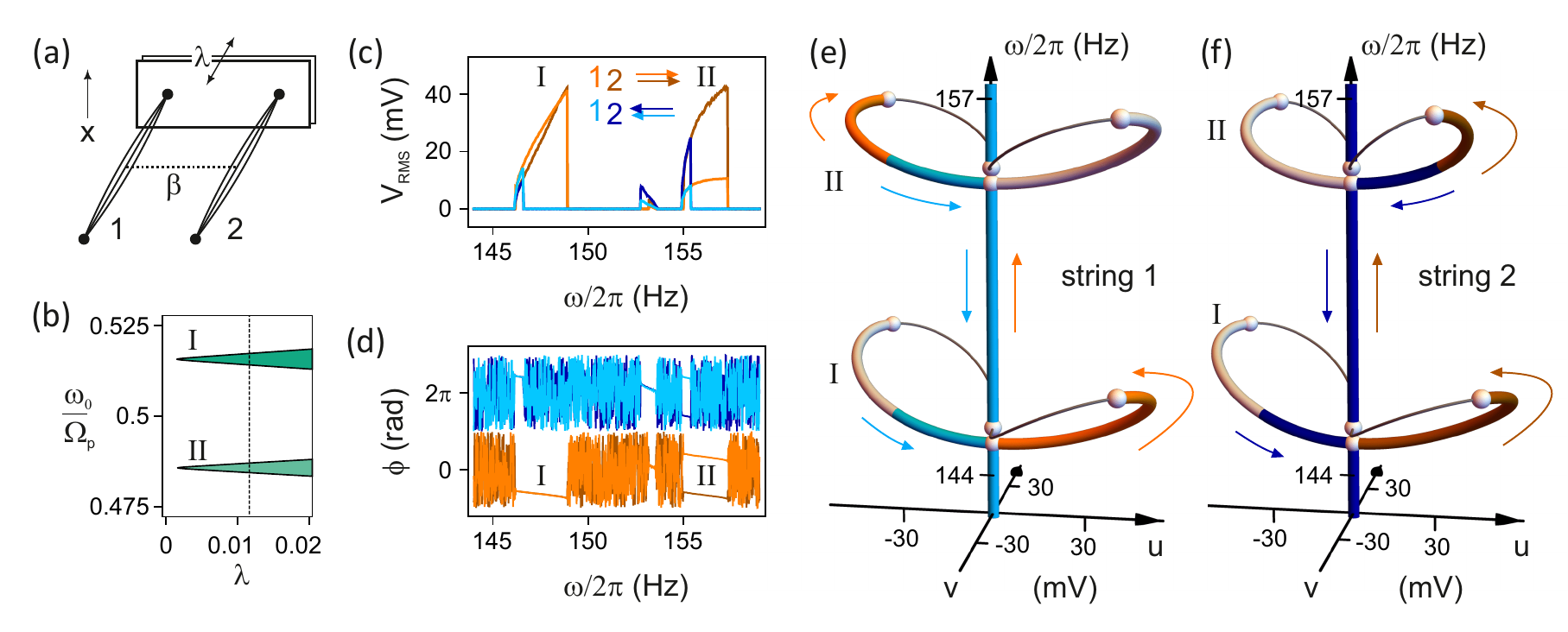}
\caption{\label{fig:figure2}  Strongly coupled oscillators: (a) Schematic setup representing two parametrically-driven  strings coupled via an additional mechanical connection. (b) Calculated normal-mode stability diagram of the symmetric and antisymmetric eigenmodes of the coupled system. (c) Measured amplitude and (d) phase of strings 1 and 2 for the upsweep (orange and brown) and  downsweep (light and dark blue) where both oscillators are parametrically driven at frequency $2\omega$.  (e)-(f) Simulated steady-state solutions of oscillators 1 and 2 in the rotating frame phase space ($u$, $v$) calculated from the slow-flow equations, cf.~Eq.~\eqref{eq:coupled_slow_7} as a function of $\omega$. The thick (thin) tubes are stable (unstable) solutions and white spheres indicate the positions of bifurcations. The stable branches corresponding to the experiment are highlighted in matching colors for up- and down-sweeps.}
\end{figure*}

We now report on an experimental demonstration of many-body TTSB  in a system of two  coupled mechanical modes. Our setup is based on the lowest transverse vibrational modes of two macroscopic strings. The strings are clamped onto a fixed frame at one end, while the other end is attached to a stiff plate that has two purposes; firstly, the plate can be driven into vibrations parallel to the string axes by an electric motor. These vibrations modulate the tension inside the strings and generate parametric pumping of both string modes. Secondly, the plate transmits vibrations between the strings, which leads to weak intrinsic coupling between the modes. In some experiments, we introduce strong mode coupling by way of a mechanical connection close to the mode antinodes, see Fig.~\ref{fig:figure2}(a).

The motion of each string is independently measured with a dedicated piezo detector embedded into one clamping point. We use a lock-in amplifier (Zurich Instruments HF2 LI) to actuate plate vibrations and to read out the electrical signals from the two piezo detectors, which are proportional to the strings' displacements. All measurements in this work were carried out in the form of frequency sweeps, where the actuation frequency $\Omega_p=  2 \omega$ and the detection frequency $\omega$ were swept slowly to capture the steady-state response of the modes.

We use weak external driving for calibration of the modes, similarly to the procedure outlined in Ref.~\cite{Leuch2016, Eichler2018}. In these experiments, the vibration amplitude is kept low and the influence of the intrinsic coupling is negligible. From the Lorentzian response of each mode, we extract typical values for $\omega_{1,2}/2\pi = 155 \pm 10$\,Hz (depending on ambient temperature) and $Q_{1,2} \sim 1200$, while we calculate the effective mass $m = 4.14\times 10^{-4}$\,kg from the geometry of the strings.  By fitting to the large-amplitude response under strong parametric pumping, we obtain the coefficients of the nonlinear potential term, $\alpha_1 = 11.93$\,mV$_{\rm{RMS}}^{-2}$s$^{-2}$ and $\alpha_2 = 6.24$\,mV$_{\rm{RMS}}^{-2}$s$^{-2}$, as well as those of the nonlinear damping, $\eta_1 = 7.1\,\mu$V$_{\rm{RMS}}^{-2}$s$^{-1}$ and $\eta_2 = 3.9\,\mu$V$_{\rm{RMS}}^{-2}$s$^{-1}$ (in the strong coupling case, we find $\eta_1 = 3.55\,\mu$V$_{\rm{RMS}}^{-2}$s$^{-1}$ and $\eta_2 = 1.95\,\mu$V$_{\rm{RMS}}^{-2}$s$^{-1}$)~\cite{supmat}. Finally, in the presence of strong coupling, we use the normal mode frequency splitting to estimate $\beta = 36.2 \pm 0.1$\,Hz.  

{\it Strong coupling} [Fig.~\ref{fig:figure2}(a)]: we first explore the regime where the two instability lobes corresponding to the symmetric and antisymmetric normal modes are well separated, see Fig.~\ref{fig:figure2}(b). In Figs.~\ref{fig:figure2}(c) and (d), we show the measured amplitudes and phases of both strings under a common parametric modulation as a function of frequency $\omega/2 \pi$, respectively. As the frequency is slowly swept upwards, both resonators oscillate with the same phase from $146$\,Hz up to $149$\,Hz. As the frequency is ramped further, the resonators are in opposing phase states from $155$\,Hz up to $157.5$\,Hz. The modes exhibit identical symmetries (s/a) when the frequency is swept downwards. These qualitative observations were consistent over many sweeps. The small peaks around $\omega/2\pi=153$\,Hz correspond to an unidentified eigenmode in the experimental setup that does not appear to affect the modes of interest.

We model the system with Eq.~\eqref{eq:coupled_slow_7} for $N=2$ using the parameters extracted from the experiment. The results of our calculations provide a simple understanding of the experimental observations: as the frequency is swept,   either the symmetric or antisymmetric normal modes undergo TTSB at their respective instability thresholds, recreating the effective single-mode TTSB discussed earlier. The coupling between the normal modes induced by nonlinearities is irrelevant in this regime as one mode is strongly off-resonant with the other. The  experimental results  are well described by  the phase-space bifurcation diagrams for each resonator plotted in Figs.~\ref{fig:figure2}(e) and (f). Despite the fact that both resonators participate in the TTSB of the symmetric or antisymmetric modes,  many-body TTSB is not observed in this strong-coupling limit as the two instability lobes do not overlap  for experimentally accessible parametric excitation  strengths. 

\begin{figure*}[t!]
\includegraphics[width = 1\textwidth]{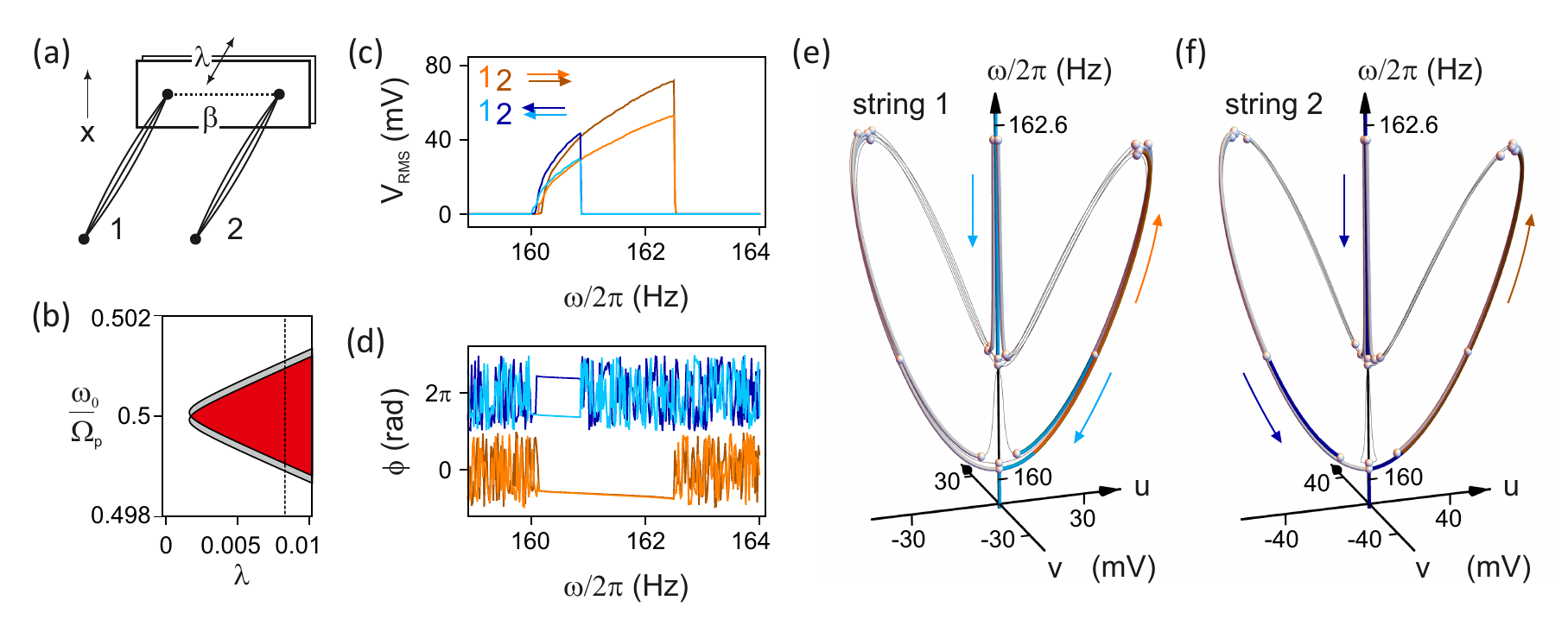}
\caption{\label{fig:figure3}  Weakly coupled oscillators: (a) Schematic setup representing two parametrically-driven  strings weakly coupled via the driving plate. (b) Normal mode stability diagram. (c) Amplitude and (b) phase of  strings 1 and 2 for the up-sweep (orange and brown) and  down-sweep (light and dark blue) where both oscillators are parametrically-driven at frequency $2 \omega$.  (c) and (d) show the simulated steady-state solutions of oscillators 1 and 2 in the  rotating-frame phase space ($u$, $v$) calculated from the slow-flow equations, cf.~Eq.~\eqref{eq:coupled_slow_7} as a function of  $\omega$. The thin  tubes are unstable  solutions  and all other colored tubes represent   stable solutions. The white spheres in these plots denote bifurcations. }
\end{figure*}

{\it Weak coupling}: next, we remove the connection between the strings and rely on the driving plate to provide weak coupling between the string modes [Fig.~\ref{fig:figure3}(a)-(b)]. The experimental data look very different in this regime [Fig.~\ref{fig:figure3}(c)-(d)]. Both strings have nearly identical  natural  frequencies (within $50$\,mHz from each other) and exhibit  hysteresis  when sweeping the frequency upwards and downwards.  The frequencies where the oscillation drops to zero (during upsweeps) or jumps to a finite amplitude (during downsweeps) are precisely the same for both resonators. The strings oscillate in phase during the upsweep and out of phase during the downsweep. All of these features were reproduced over many sweeps. 

The theoretical model corresponds to normal modes that are split by a very small coupling $\beta$, such that their instability lobes overlap strongly [Fig.~\ref{fig:figure3}(b)]. Since both normal modes exhibit TTSB and are weakly coupled by nonlinearities, we witness the realization of two-body TTSB. As before, the experimental results for the amplitude and phase are consistently explained by the weak coupling bifurcation diagram for both strings shown in Figs.~\ref{fig:figure3}(e)-(f).  In comparison with the strong coupling scenario of Figs.~\ref{fig:figure2}(e)-(f), the weakly coupled system exhibits richer behavior. The selection of symmetric and antisymmetric solutions as a function of the sweeping direction may be explained in terms of the phase response of a  linear resonator to a periodic external force. Below its natural frequency, a harmonic resonator oscillates with almost no phase lag in response to an external force. As the two string modes drive each other, they prefer to move in phase. In contrast, since the harmonic resonator response has a phase lag of $\sim \pi$ above the natural frequency, the string modes preferably oscillate out of phase during the downsweep. This many-body TTSB state is stable against small detunings $\omega_1 \ne \omega_2$ and robust to noise (as seen in the experiment). Increasing noise levels are expected to preserve the underlying TTSB, but to induce transitions between the different stable solutions.

Coupled parametric resonators provide  the simplest platform to realize macroscopic states with robust discrete time-translation symmetry breaking. Period-doubling bifurcations in stable steady-states provide a rich space of solutions that manifest such many-body phenomena. This can be readily generalized to the quantum realm, where the bifurcations physics is replaced by dissipative first- and second-order phase transitions~\cite{ Minganti2016,Bartolo2016,Elliott2016,Heugel2019}. Furthermore, higher-period TTSB can also be realized in these systems through a judicious choice of modulated nonlinearities~\cite{Zhang2017A}. In the weak coupling limit, the classical network can be viewed as an Ising machine that simulates complex problems, where the system parameters can be tuned to engineer desired `spin configurations' of the Ising-like phase states. The analogous quantum network comprising dissipative Kerr parametric resonators is expected to  manifest an equivalent TTSB phase \cite{Savona2017,Rota2018,Heugel2019, Minganti2016,Bartolo2016,Elliott2016}. Such networks have been proposed as quantum annealers~\cite{Puri2017}, and following this work can now be used as quantum simulators of many-body time crystals.


\vspace{5mm}
\textbf{Acknowledgments} 
This work received financial support from the Swiss National Science Foundation through grants (CRSII5\_177198/1) and (PP00P2\_163818). 


\vspace*{1cm}

\clearpage
\newpage
\newpage

\newpage

\cleardoublepage
\setcounter{figure}{0}
\renewcommand{\figurename}{Supplementary Material Figure}

\begin{center}
\textbf{\normalsize Supplemental Material for}\\
\vspace{3mm}
\textbf{\large Classical many-body time crystals}
\vspace{4mm}

{ Toni L. Heugel$^{1,\ast}$, Matthias Oscity$^{1,2,\ast}$, Alexander Eichler$^{3}$, Oded Zilberberg$^{1}$, and R. Chitra$^{1}$}\\
\vspace{1mm}
\textit{\small $^{1}$Institute for Theoretical Physics, ETH Z\"{u}rich, Wolfgang-Pauli-Stra\ss e 27, 8093 Z\"urich, Switzerland.\\
$^{2}$Fachhochschule Nordwestschweiz FHNW, Klosterzelgstrasse 2, CH-5210 Windisch, Switzerland.\\
$^{3}$Institute for Solid State Physics, ETH Z\"{u}rich, Wolfgang-Pauli-Stra\ss e 27, 8093 Z\"urich, Switzerland.\\
$^\ast$ These authors contributed equally.
} 

\vspace{5mm}
\end{center}

\setcounter{equation}{0}
\setcounter{section}{0}
\setcounter{figure}{0}
\setcounter{table}{0}
\setcounter{page}{1}
\makeatletter

\renewcommand{\thefigure}{S\arabic{figure}}

\setcounter{enumi}{1}
\renewcommand{\theequation}{S\Roman{enumi}.\arabic{equation}}
\renewcommand{\thesection}{\Roman{section}}

\section{Derivation of $\lambda_{th}$}
We present here the derivation of the many-body parametric driving threshold amplitude for $N$  resonators that are equally coupled to one another. For the prupose of this calculation, it suffices to consider $N$ coupled \emph{linear} resonators.   
The slow-flow equation describing this system (cf. Eq.(2) in the paper with coupling $\beta_{ij}=\beta$ for all $i\ne j$) is given by:
\begin{equation}\label{eq:sf}
\dot{\bm{X}} = A \bm{X},
\end{equation}
where the matrix $A$ is given by:
\begin{align}
A &= \begin{pmatrix}
a & b & \cdots &b\\
b&a & \ddots  & \vdots \\
\vdots & \ddots &\ddots &b \\
b&\cdots&b&a
\end{pmatrix}.
\end{align}
and the  individual matrix entries $a$ and $b$ are given by:
\begin{align}
a &= \left(\begin{array}{cc} -\frac{\gamma}{2}  & -\frac{1}{4 \omega} \left(\lambda \omega_0^2 +2(\omega_0^2 - \omega^2)\right) \\  
-\frac{1}{4 \omega} \left(\lambda \omega_0^2 -2(\omega_0^2 - \omega^2)\right) & -\frac{\gamma}{2} \end{array}\right), \\
b &=  \left(\begin{array}{cc} 
0 &  \frac{\beta}{2 \omega} \\  
- \frac{\beta}{2 \omega} & 0 
\end{array}\right)\,.
\end{align}

The dynamics of the linear system can be deduced  by decomposing the initial state $\bm{X}$ into the eigenvectors of $A$. The time evolution of each eigenvector is then determined by $e^{\Lambda t}$, where $\Lambda$ is the corresponding eigenvalue. ${\rm Im}\Lambda \ne 0 $ imposes an oscillatory behavior whose envelope decreases exponentially for  ${\rm Re}\Lambda < 0 $ and  increases exponentially for  ${\rm Re}\Lambda > 0 $.
 To evaluate these eigenvalues and eigenvectors, it is useful to rewrite the matrix $A$ as:
\begin{equation}
A = Id_N \otimes a +  \left(\begin{array}{cccc} 
0  &  1 & \cdots & 1\\
1 & \ddots &\ddots &\vdots\\
\vdots & \ddots & \ddots &1\\
1 &\cdots & 1& 0
\end{array}\right) \otimes b = Id_N \otimes a+ M_N\otimes b\,.
\end{equation}
Based on the structure of $A$,  the eigenvectors obey  the ansatz $w_m = r_m \otimes s_m$, where $r_m $ are the $N$ dimensional eigenvectors of $M_N$ with eigenvalue $\rho_m$ and $s_m$ are the 2-dimensional eigenvectors of $a+\rho_m b$ with eigenvalues $\sigma_m$. Since $M_N$ has $N$ eigenvectors, $r_m$, with $2$ corresponding $s_m$, this ansatz describes all the $2N$ eigenvectors and eigenvalues of the matrix $A$. The 2-vector $s_m$ describes the amplitude and momentum $(u_m,v_m)$ in a particular mode's phase-space and $r_m$ generically describes the relative amplitudes and the phase configuration, e.g., $r_m = (1,-1)$ means that the two oscillators have opposite phases. We can readily show that this ansatz is indeed an eigenvector of $A$:
\begin{align}
A w_m &=  Id_N r_m \otimes a s_m + M_N r_m\otimes b s_m \\
&= r_m \otimes (a + \rho_m b) s_m\nonumber\\
&= \sigma_m w_m\,.\nonumber
\end{align}

Since $M_N$ has a simple structure,  we see that  the eigenvectors $\{r_m\}$ take the form  $r_0=(1,1,\cdots)$ with eigenvalue $\rho_0 = N-1$ and  $r_m = (0, \cdots 0, 1, -1, 0 \cdots)^T$, where  the $+1$ is the $m^{th}$ entry, are eigenvectors of $M_N$ with eigenvalues $\rho_m = -1$ ($m \in \mathbb{N}$, $0<m \leq N - 1$).  Note that the eigenvectors $r_m$ effectively determine the normal mode transformations of the problem.
Next, we evaluate  the  eigenvectors $s_m$ and eigenvalues $\sigma_m$ of 
\begin{align}
a + \rho_m b = \left(\begin{array}{cc} 
a_1 & a_2 -a_3  \\
a_2 + a_3 & a_1
\end{array}\right)\,,
\end{align}
where $a_1 = -\frac{\gamma }{2}$, $a_2=-\frac{ \lambda  \omega_0^2}{4 \omega }$ and $a_3= \frac{ \left(\omega_0^2-\omega ^2\right)}{2 \omega } -\rho_m \frac{\beta }{2 \omega }$.
These are given by, 
\begin{align}
\sigma_{m,\pm} &= a_1 \pm \sqrt{a_2^2 - a_3^2}\,,\\
s_{m,\pm} &=  \left(\begin{array}{cc} 
\pm \sqrt{a_2 -a_3}  \\
\sqrt{a_2 + a_3}
\end{array}\right)\,.
\end{align}
To summarize, the  $2N$ eigenvectors of the matrix $A$ are given by
\begin{align}\label{eq:wc}
w_{m,\pm} &= r_{m} \otimes s_{m,\pm}\,,
\end{align}
with corresponding eigenvalues  $\sigma_{m,\pm}$ and $0\le m \leq N-1$.
 
If ${\rm Re}\sigma_{m,\pm} >0 $, the corresponding $w_{\pm,m}$ grows exponentially  indicating a parametric instability.  We  obtain  the parametric driving threshold  $\lambda_{th, m}$ for this instability by imposing the condition:
\begin{align}\label{eq:lamth}
\sigma_{m,+} = a_1 + \sqrt{a_2^2 - a_3^2} &= 0\,.
\end{align}
Note that we have $a_1<0$, whereas $\sqrt{a_2^2 - a_3^2}$ can be either real-valued and positive or complex-valued.  Solving Eq.~\ref{eq:lamth}, we obtain
\begin{align}
\lambda_{th, m} = \frac{4\omega}{\omega_0 ^2}\sqrt{a1^2 + a_3^2} &= \frac{4\omega}{\omega_0 ^2}\sqrt{\frac{\gamma^2}{4} +  \left( \frac{\omega^2 - \omega_0^2}{2\omega } + \rho_m \frac{\beta}{2 \omega}\right)^2}\,.
\end{align}
For identical oscillators, we see that there are primarily two instability thresholds corresponding to 
(i) the instability of the symmetric normal mode, $w_{0,+}$, and (ii) to the instability of all other normal modes: $w_{m,+}$ including the antisymmetric mode.

\section{Calibration measurements}

\begin{figure}[ht!]
\includegraphics[width=\columnwidth]{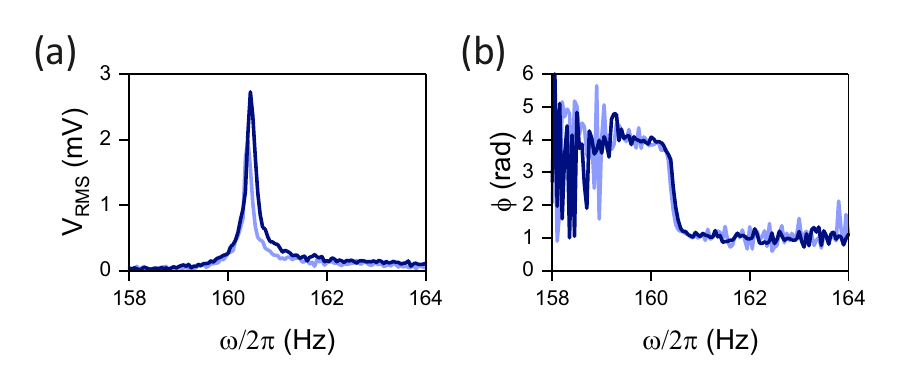}
\caption{\label{fig:figureS1} (a) Amplitude and (b) phase response of the two resonators in the linear regime. We use weak external driving, no parametric drive, and weak coupling to observe a Lorentzian response. Light and dark blue correspond to resonator 1 and 2, respectively. These measurements are taken immediately before the nonlinear parametric measurements shown in Fig. 3 of the main text to ensure that the two modes are degenerate in frequency.}
\end{figure}

\begin{figure}[ht!]
\includegraphics[width=\columnwidth]{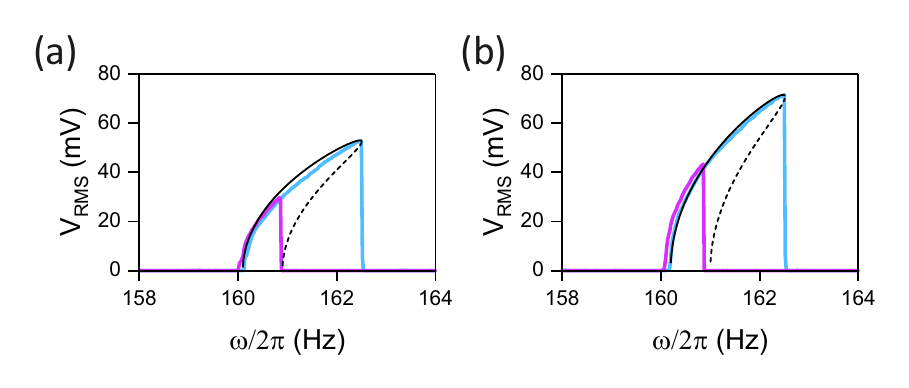}
\caption{\label{fig:figureS2} (a) Amplitude response of resonator 1 and (b) resonator 2 with weak coupling and strong parametric driving. Blue and magenta lines correspond to sweeps with increasing and decreasing frequency, respectively. These are the same data as shown in Fig. 3c of the main text. Solid and dashed black lines are stable and unstable theory solutions, respectively. From fitting these solutions to the measured data, we retrieve the values of $\alpha_{1,2}$ and $\eta_{1,2}$ stated in the main text. Note that in the strong coupling case we find that the nonlinear damping decreases, as determined from the frequency at which the stable and unstable solutions merge.}
\end{figure}

In Fig.~\ref{fig:figureS1}, we present test measurements that we have performed to ensure that the weakly coupled strings were degenerate in frequency. On timescales of hours, thermal drift sometimes caused detuning between the strings, which we balanced by adjusting the tension of the strings separately. In Fig.~\ref{fig:figureS2}, we show the fits used to extract the nonlinear coefficients of the two weakly coupled strings. Please refer to Ref.~[23] of the main text for details regarding the model of a nonlinear parametric oscillator.
 
\end{document}